# Electrical and optical studies of GaMnAs/GaAs(0 0 1) thin films grown by molecular beam epitaxy

J.F. Xu, S.W. Liu, M. Xiao, P.M. Thibado

Department of Physics, University of Arkansas, Fayetteville, Arkansas 72701, AR, USA

Abstract
GaMnAs/GaAs films were grown via molecular beam epitaxy using both low and high substrate temperatures. The films were investigated using Hall effect and photoluminescence (PL) measurements from 8 to 300 K. The carrier concentrations in the samples grown at a low substrate temperature are greater than those in the samples grown at a high substrate temperature. The PL spectra show a GaAs exciton peak, a peak involving a carbon acceptor, a substitutional Mn acceptor-related peak and an optical phonon-related peak.

1. Introduction

Recently, the p-type dilute magnetic semiconductor (DMS) Mn-doped GaAs (GaMnAs) has been attracting much attention. GaMnAs can be described qualitatively as a random alloy in which Mn substitutes at Ga sites and takes the dual role of acceptor and local magnetic moment. The combination of ferromagnetism with the versatile semiconducting properties makes it promising for future spintronics applications. In addition, GaMnAs can be grown using several techniques, including molecular beam epitaxy (MBE) [1,2], liquid phase epitaxy (LPE) [3] and ion implantation [4,5]. The realization of spintronic devices requires a substantial increase in the ferromagnetic transition temperature ($T_c$) as the highest in GaMnAs is around 100 K. Because of the low solubility of Mn in GaAs, it is thought that most Mn ions form in the DMS as defects, such as interstitials. These defects compensate substitutional Mn acceptors ($Mn_{Ga}$) and, in turn, lead to a suppression of the ferromagnetic transition temperature [6]. It has been demonstrated that $T_c$ exceeding 150K can be achieved by low-temperature annealing in films thinner than _60nm [7,8]. Post-growth annealing is thought to enable Mn interstitials to diffuse to the surface thereby maximizing the concentration of uncompensated $Mn_{Ga}$ and thus $T_c$. These results show this material is highly sensitive to the growth and post-growth conditions. A small change in alloying compositions or microstructures can induce significantly different physical properties. Thus, it is important to study the electrical and optical properties in the GaMnAs system as a function of growth parameters.

In this paper, we discuss DMS GaMnAs thin films that were grown at high and low substrate temperatures using MBE. These samples were investigated using Hall effect and PL measurement from 8 to 300 K. Carrier concentration and Hall mobility were evaluated, and the energy of the Mn acceptor level was determined from the photoluminescence (PL) data.

2. Experimental details

Samples were prepared in an ultrahigh vacuum (~$2\times10^{-10}$ Torr) MBE growth chamber (Riber 32) that includes Ga and Mn effusion cells together with a two-zone As valved-cracker cell. The MBE chamber is also equipped with a reflection high-energy electron diffraction (RHEED) system. Commercially available, ''epi-ready,'' semi-insulating 2 inch diameter GaAs(0 0 1) wafers were cleaved into quarters. One quarter was mounted on a 2 inch diameter standard MBE molybdenum block using indium as solder. The substrate was then loaded into the load–lock chamber without any chemical cleaning. Next, the substrate was transferred to the heating stage inside the MBE chamber and the MBE chamber was cooled down using liquid nitrogen. The substrate was heated to 580 C while exposing the surface to $As_4$ to remove the surface oxide layer. A thin buffer layer of GaAs was grown on the substrate for 1–5 min. During this time, RHEED oscillations were used to determine that the growth rate of the GaAs was 780 nm/h. Next, the substrate temperature ($T_s$) was set to the desired growth temperature of either 580 or 250 C. GaMnAs films were then grown for 1 h while RHEED was used to monitor the surface reconstruction during and following the growth. After growth, the sample was cooled down and removed from the UHV system. Samples were cleaved into multiple smaller pieces (5mm×5 mm) for characterization measurements.

The carrier concentration and Hall mobility were estimated from low-field (H = 0.2 T) Hall effect measurements at room temperature and 77 K. The van der Pauw method was used on the 5mm×5mm square samples. Ohmic contacts were formed by soldering point-size indium contacts at each corner of the sample. The PL measurements were performed in a variable temperature (8–300 K) closed-cycle helium cryostat. The 532 nm line from a double Nd:YAG laser was used for continuous wave PL excitation. The PL signal from the sample was dispersed by a monochromator and detected by a liquid nitrogen-cooled charged-coupled device (CCD).

3. Results

The relation between the carrier concentration and Mn cell temperature used to grow the samples is shown in Fig. 1. The majority carrier type in all of the samples was found to be holes. The hole concentrations were measured at both room and liquid nitrogen temperatures. The low substrate temperature ($T_s$) sample data are shown in Fig. 1(a). As the Mn cell temperature is increased the hole concentration increases, reaches a maximum value, but then it decreases. The concentration peak positions lie at 800 and 850 C for room temperature and 77K measurements, respectively. The hole concentration peaks in the $10^{19}$ cm$^{-3}$ range, but shows a very large dynamic range. What is surprising is that the hole concentration goes down even though more Mn is added to the sample at the higher Mn cell temperatures. For high substrate temperature samples, the carrier concentration mostly increases with Mn cell temperature as seen in Fig. 1(b). Here, however, the concentration peaks in the $10^{18}$ cm$^{-3}$ range, or about a factor of 10 less than the low $T_s$ samples.

The dependence of the hole mobility on the Mn cell temperature for both high and low substrate temperature ($T_s$) grown samples is shown in Fig. 2. For the low $T_s$ samples, the hole mobility measured at room temperature and 77K are nearly identical as seen in Fig. 2(a). Both values are rather low and do not change much with Mn cell temperature. However, the hole mobility does increase at a Mn cell temperature of 900 C compared to other Mn cell temperatures. It is surprising that the hole mobility in the low $T_s$ samples at 77K is not higher than that at room temperature, even though the carrier concentration is lower (see Fig. 1). For the high $T_s$ samples, the hole mobility at room temperature is lower than at 77K as seen in Fig. 2(b). At room temperature, the hole mobility basically does not change with Mn cell temperature. At 77 K, the hole mobility is much higher and drops with increasing Mn cell temperature. Interestingly, the high $T_s$ sample's mobilities are 10–100 times greater than the low $T_s$ samples.

The PL spectra of the GaMnAs films grown using a Mn cell temperature of 800 C but different substrate temperatures are shown in Fig. 3. The spectra for the two samples were taken at 8K and are similar to each other with the corresponding peak positions, but the peaks in high $T_s$ sample (lower curve) are much stronger. Several luminescence peaks can be identified, namely, the GaAs exciton line at 1.5057 eV, an impurity-induced recombination peak at 1.4865 eV and some deeper bands near 1.4515, 1.4041, and 1.3686 eV. In low-$T_s$ sample, the 1.4515 eV peak is present but weaker.

4. Discussion

It is interesting to notice in Fig. 1(a) that the hole concentration in low-$T_s$ samples increases with Mn cell temperature at first but then decreases. We expect an increase with Mn cell temperature because the hole concentration comes from Mn and is proportional to the Mn flux rate. The drop in the hole concentration with the Mn cell continues to increase, but this is hard to understand. It is known that Mn does have a low solid solubility limit (~$10^{19}$ cm$^{-3}$) in GaAs. We believe the hole concentration drops at higher Mn cell temperatures, because the Mn concentration begins to exceed its solid solubility limit, in GaAs. As more Mn is added above the solid solubility limit some defects will likely form, such as Mn-rich precipitates and Mn interstitials. These defects can lead to a decrease in the hole concentration due to charge compensation factors.

In high-$T_s$ samples shown in Fig. 1(b), we see a more normal relationship. Here, the Mn hole concentration increases as the amount of Mn increases. However, the unusual feature here is the low carrier concentration. The carrier concentration is always below the solid solubility limit (~$10^{19}$ cm$^{-3}$). We believe that the Mn deposited in a given plane of the GaAs film does not stay in that plane, but floats up along the growing films surface. This can happen when the substrate temperature is higher, and this would result in the lower carrier concentration.

By carefully analyzing the hole mobility data shown in Fig. 2, we can learn more about the physical nature of these crystals. The main striking discrepancy for the mobility data is that the low substrate temperature mobilities are dramatically lower than the high substrate mobilities. In fact, the high substrate temperature mobilities are similar to what one would find in a pure GaAs single crystal with few defects and a low number of carriers. We believe the low substrate temperature mobility numbers are a reflection of the poor crystal quality of the thin films. It is known that growing low-temperature GaAs does result in a much lower crystal quality and that the density of those films is lowered to about 80–90% of the high substrate temperature grown materials.

We can learn more about these samples from the PL data shown in Fig. 3. Three of the PL peaks can be easily identified from the literature. According to the location and width, the 1.5057 eV peak should be the GaAs exciton radiative transition. The stronger GaAs exciton radiative transition peak in high-$T_s$ sample indicates that it has a much higher structural quality. The peak at 1.4865 eV is also well known. It is the donor–acceptor pair transition involving a carbon at an As site [9]. The emission band around 1.3686 eV is thought to be the longitudinal–optical (LO) phonon mode in GaMnAs [10].

The most interesting PL peak is at 1.4041 eV. This peak is associated with the transition from the conduction band to the Mn acceptor level. This peak is a direct confirmation that Mn atoms are occupying Ga sites [10,11]. In fact, this Mn-related peak gives rise to an acceptor-binding energy of 101.6 meV above the valence band edge, which is in good agreement with values obtained by many different methods [12,13]. This peak indicates at least some of Mn is indeed incorporating in the desired substitutional Ga sites. It seems the peak in high-$T_s$ sample is even stronger. The stronger Mn substitution emission peak is surprising as the hole concentration in high-$T_s$ samples is so low [see Fig. 1(b)]. However, the background level in the low-$T_s$ material makes a quantitative comparison difficult.

The PL emission peak at 1.4515 eV has never been reported before in GaMnAs or GaAs. We can see this peak in both low-and high-$T_s$ samples although it is stronger in the high-$T_s$ one. This peak may be a Mn-related emission; however, further studies are needed.

It is important to compare and contrast our results with previous studies of Mn-doped GaAs. For the GaMnAs thin films prepared with LPE [3,10,11] or ion implantation [5,13] most of the characterization is done using PL. For these growths they observe Mn acceptor-transition peaks but only for the low-Mn doping, not high-Mn doping like our studies. We believe this indicates that the crystalline quality is higher in our MBE-grown samples. For the GaMnAs thin films prepared by MBE [1], similar high and low substrate temperature growths have been carried out and the samples have been characterized using Hall measurements but not PL. Their Hall carrier concentration results are similar to ours for high substrate temperature samples, but different from ours for low substrate temperature samples. In particular, our hole concentration increase by two orders of magnitude and then decreases by two orders of magnitude. The drop is likely due to some compensation mechanism. In addition, with our growth recipe the hole mobilities are about 10 times higher, which would again point to a higher crystalline quality.

5. Conclusions

The Mn-doped GaAs (GaMnAs) thin films were grown on semi-insulating GaAs(0 0 1) substrates using MBE with both high and low substrate temperatures. The carrier concentration in low-$T_s$ samples increases with Mn cell temperature first, but then it decreases. This is likely due to the solubility limit of Mn in GaAs being reached. Mobility measurements show that the crystalline quality of the low-$T_s$ samples is much poorer than those grown at high substrate temperatures. PL transitions involving Mn acceptors were clearly identified in both low-and high-$T_s$ samples. This means that some of the Mn is going into Ga sites regardless of the substrate temperature used.

Acknowledgment

The authors would like to acknowledge the support for this work from National Science Foundation under grant number DMR–0405036.

Fig. 1. Hole carrier concentration in the Mn-doped GaAs films grown at different substrate temperatures ($T_s$) as a function of the Mn cell temperature. Measurements were carried out at both room temperature (shaded square) and 77K (shaded triangle): (a) data for samples grown using a low substrate temperature ($T_s$) equal to 250 C and (b) data for samples grown using a high substrate temperature ($T_s$) equal to 580 C.

Fig. 2. Hole carrier mobility in the Mn-doped GaAs films grown at different substrate temperatures ($T_s$) as a function of the Mn cell temperature. Measurements were carried out at both room temperature (shaded square) and 77K (shaded triangle): (a) data for samples grown using a low substrate temperature ($T_s$) equal to 250 C and (b) data for samples grown using a high substrate temperature ($T_s$) equal to 580 C.

Fig. 3. PL spectra acquired at 8K for the Mn-doped GaAs films grown using a Mn cell temperature of 800 C and two different substrate temperatures ($T_s$): $T_s$ = 250 C (upper curve) and $T_s$ = 580 C (lower curve).

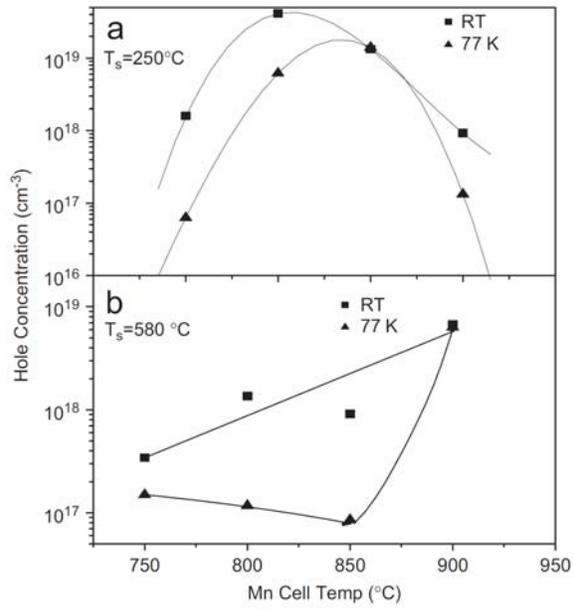

Figure 1.

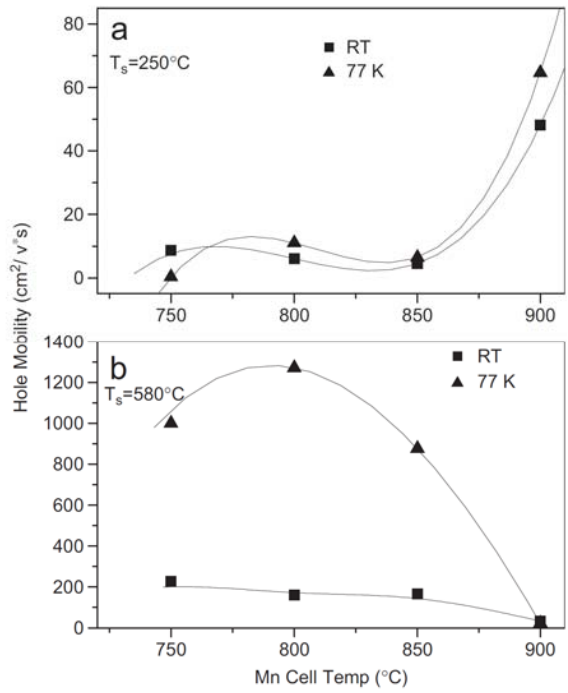

Figure 2.

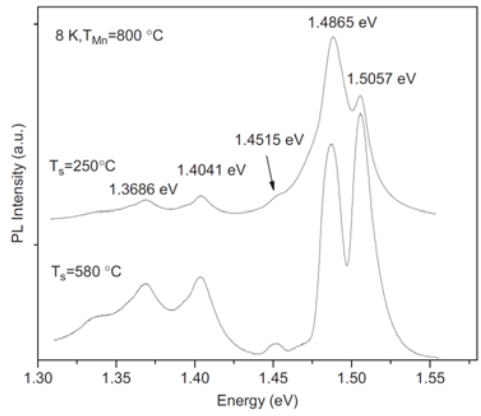

Figure 3.